\documentstyle[aps,epsf,array]{revtex}

\newcommand{\qp}{{\bf q}_{\perp}}

\newcommand{\qpm}{q_{\perp}}

\newcommand{\rp}{{\bf r}_{\perp}}

\newcommand{\spq}{\langle u({\bf q_{\perp}}) ^2 \rangle }

\newcommand{\nabper}{\nabla_{\perp}}

\begin{document}

\draft


\title{Active Membrane Fluctuations Studied by Micropipet Aspiration}

\author{J-B. Manneville$^{1}$,
P. Bassereau$^{1}$\footnote{To whom correspondence should be addressed
(e-mail: Patricia.Bassereau@curie.fr)}, S. Ramaswamy$^{2}$, and J. Prost$^{1}$}

\address{$^{1}$UMR CNRS-Curie 168, Institut Curie
26 Rue d'Ulm, 75248 Paris Cedex 05, France \\
$^{2}$Center for Condensed Matter Theory, Department of Physics, Indian 
Institute of Science, Bangalore 560 012, India}

\maketitle

\begin{abstract}

We present a detailed analysis of the micropipet experiments recently reported
in J-B. Manneville {\it et al.}, Phys. Rev. Lett. {\bf 82}, 4356--4359 (1999), 
including a derivation of the expected behaviour
of the membrane tension as a function of the areal strain in the case of
an active membrane, {\it i.e.}, containing a nonequilibrium noise
source. We give a general expression, which takes into account the effect of 
active centers both directly on the membrane, and on the embedding fluid 
dynamics, keeping track of the coupling between the density of active centers
and the membrane curvature.
The data of the micropipet experiments are well reproduced by the new 
expressions. In particular, we show that a natural choice of the parameters 
quantifying the strength of the active noise explains both the large
amplitude of the observed effects and its remarkable insensitivity to the 
active-center density in the investigated range.
\end{abstract}

\pacs{87.22.Bt Membrane and subcellular physics and structure -
82.65.Dp Thermodynamics of surfaces and interfaces}

\section{Introduction}

Biological membranes are made up of a complex mixture of lipids and
proteins. The lipid molecules form a bilayer 
structure 
which separates the cytoplasm of the cell from the outside.
In addition to this structural role, the membrane also participates
in a number of the living cell functions
\cite{ReviewsMbBio}, mostly performed by proteins embedded inside the
lipid bilayer,
such as solute transport via ion channels
or pumps, cell locomotion and adhesion, membrane transport through
exocytic and endocytic pathways, signal transduction\ldots\
Consequently, from the statistical physics point of view, biological membranes
are strongly out of thermodynamic equilibrium,
whereas most studies on membranes reported in the physics
literature have been done
at thermodynamic equilibrium  \cite{ReviewsMbPhy}.
In order to achieve a more complete physical description of biological
membranes, this nonequilibrium aspect
clearly has to be included.
The field of membrane shape fluctuations is a good test case in which  
to examine the relevance of nonequilibrium effects.  
At thermodynamic equilibrium,
the membrane shape fluctuates because of thermal noise, {\it i.e.}, the Brownian 
motion of the bilayer.  
Such membrane will be called a \lq passive\rq\ membrane in this paper. 
If a nonequilibrium noise source is superimposed to 
thermal noise, due for instance
to the activity of membrane proteins, then the membrane is no
longer at thermodynamic equilibrium. In that case, the membrane
will be called \lq active\rq.

Recently, micropipet experiments on fluctuating giant vesicles containing
bacteriorhodopsin (BR) reconstituted in the lipid bilayer have shown that
the light-driven proton pumping
activity of BR induces an amplification of the membrane shape
fluctuations \cite{JB}. In these experiments,
a larger excess area could be pulled out
by micropipet aspiration when the proteins were activated.
The results were qualitatively interpreted in terms of 
an increase of the effective membrane temperature and were not
directly compared to theoretical predictions. 
In the present article,
we give details about the experimental procedure
(section \ref{expdetails}) and develop a
theoretical framework to analyze quantitatively the micropipet
experiments
(sections \ref{theor} and  \ref{exptheor}).
According to theory, a qualitatively new fluctuation spectrum is expected in the
presence of a nonequilibrium noise source \cite{JP1,JP2,JP3}.
These earlier theories introduced the nonequilibrium activity in the membrane
conformation equation only. This restrictive choice was made because the
nonequilibrium force-density arising from the active proteins, when included 
in the Stokes equation for the solvent velocity field, altered the membrane
fluctuation spectrum through terms which were subdominant at small wavenumber.  

We show here that those nominally subdominant terms provide the most important  
contribution in the experimentally relevant range.  
This unexpected behavior is due to the 
very small value of the permeation coefficient. With this implementation, theory 
and experiment are brought in agreement (section \ref{exptheor}). Even the absence of 
sensivity of the experimental data on the active center density appears as a
natural consequence of the developed theory.

\section{Experimental procedure and results}
\label{expdetails}

\subsection{Bacteriorhodopsin}

\begin{figure}
\epsfxsize=10cm
\centerline{\epsfbox{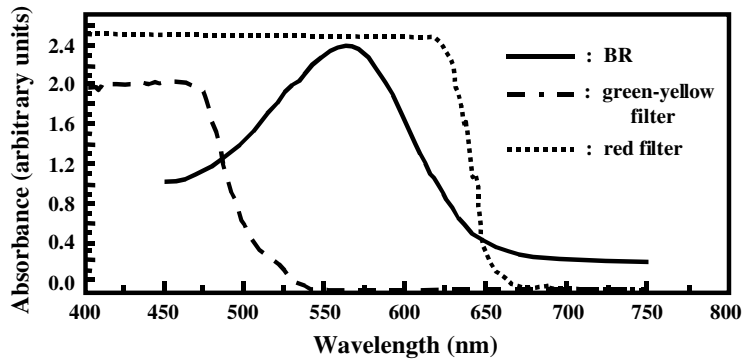}}
\caption[]{\label{brabs}
Absorption spectrum of the purple membrane suspensions used
in the experiments (solid line) superimposed with the absorption
spectra of the high-pass green-yellow filter (dashed line) and the
high-pass red filter (dotted line).
}
\end{figure}

The bacteriorhodopsin (BR) is a $27 \,$ kDa protein
\cite{ReviewsBR1} purified from the so-called purple membrane
of the halophilic bacteria {\it
Halobacterium salinarum} \cite{ReviewsBR2}.
Its structure is known at the atomic level with high resolution \cite{structBR}.
The BR absorption spectrum shows a maximum in the green-yellow wavelength
around $566$ nm (Fig. \ref{brabs}).

When illuminated with green-yellow light, proton pumping
is activated through a photocycle involving several distinct
photointermediates \cite{ReviewsBR3}. The total duration of the
photocycle is $\tau \simeq 5$ ms. Structural changes of the BR
during the photocycle have been investigated to elucidate the
translocation pathway of the proton across the protein.
The pumping mechanism has been recently completely elucidated, so that BR 
is to date the best understood ion pump \cite{pumpBR}.
The proton pumping activity has been extensively studied
in reconstituted systems, mostly in large
unilamellar vesicles ($0.1-1 \,
\mathrm{\mu m}$ in diameter) \cite{petitesves}.

\subsection{Giant vesicle formation}

In reference \cite{JB}, the electroformation technique of giant
unilamellar vesicles ($10-100 \, \mathrm{\mu m}$ in diameter) \cite{Angelova},
modified according to \cite{Daniel} 
for BR incorporation in the lipid
bilayer, was used to grow giant vesicles from a
mixed lipid/protein dried film.
The phospholipid EPC (Egg Phosphatidylcholine; Avanti Polar Lipids, Alabaster, Alabama, USA)
is a mixture of lipids with different chain
lengths and degrees of saturation
and is known to be adequate for BR incorporation
\cite{Tocanne}.
EPC ($0.5$ mg/ml) was first resuspended in diethyl ether.
Concentrated BR ($18$ mg/ml) was then added
at a molar ratio of $80$ lipid molecules per BR molecule. The mixture was sonicated at $0
\, \mathrm{^o C}$ for $30$ seconds and a few microliters were deposited
on ITO (Indium Tin Oxide) treated glass slides at $4\, \mathrm{^o C}$.
The protein/lipid film was dried under vacuum overnight. A vesicle
electroformation chamber was formed by assembling and sealing with wax
(Sigillum wax; Vitrex, Copenhagen, Denmark) two ITO slides separated by $1$ mm Teflon
spacers.
The film was hydrated by injecting a  $50$ mM sucrose solution
in the chamber.
An electric field ($1.5$ V AC) was applied across the chamber by connecting
the ITO slides to copper electrodes. Giant vesicles were obtained in
about two hours and transferred in a micromanipulation chamber filled
with $50$ mM glucose to enhance the optical contrast between the
inside and the outside of the vesicles. 
Sodium azide ($1$ mM) was first added to the sucrose and glucose 
solutions to avoid bacteria proliferation.
In some experiments,
respectively $16$ and $25 \, \%$ (w/w) glycerol was added to both the
internal and external solutions in order to increase the viscosity to
respectively $1.5 \eta_w$ and $2 \eta_w$, where $\eta_w$ is the
viscosity of water.

\begin{figure}
\epsfxsize=12cm
\centerline{\epsfbox{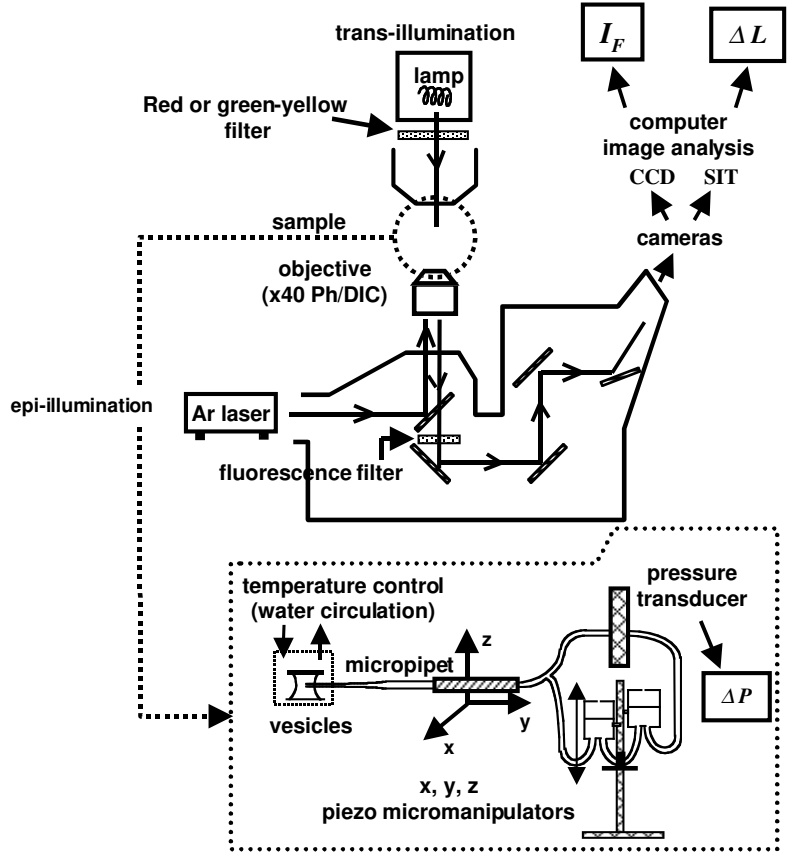}}
\caption[]{\label{schematic}
Experimental set-up designed for the micropipet experiments.
}
\end{figure}

BR incorporation was checked
by fluorescent labelling of BR with FITC (Fluorescein Isothiocyanate,
F-143; Molecular Probes, Eugene, Oregon, USA) following a published
protocol \cite{Fluo}. Excitation of the FITC was performed at $488$ nm  with an argon
laser (Spectra
Physics, Les Ulis, France) through the epi-illumination pathway of an inverted
microscope (Axiovert $135$; Zeiss, Oberkochen, Germany).
The fluorescence images of the vesicles were acquired
by a low light level SIT
(Silicon Intensified Target) camera (LH4036; Lhesa, France) (see Fig. \ref{schematic}).
The fluorescence intensity $I_F = (I_{ves}-I_{bgd})/I_{bgd}$,
where $I_{ves}$ is the fluorescence intensity of the vesicle contour
and $I_{bgd}$ is the background intensity, was
measured by computer image analysis using a
C++ custom software running on a Pentium 200-based computer
with a Meteor frame grabber (Matrox, Rungis, France).
The BR was activated by illumination through a high-pass ($500$ nm) green-yellow filter
located in the trans-illumination pathway (see Figs. \ref{brabs} and 
\ref{schematic}),
to avoid the non-pumping branched photocycle initiated if
the M intermediate absorbs at $440$ nm
\cite{ReviewsBR3}. To image the vesicles without activating the BR, a
high-pass ($650$ nm) red filter was substituted for the green-yellow
filter (see Figs. \ref{brabs} and \ref{schematic}).  
The illumination power was in the same range as that known to fully
activate BR reconstituted in
large unilamellar vesicles ($10^3 \, \mathrm{W/m^2}$)  \cite{petitesves}.
To correct for the different bandwidths of the green-yellow and
red filters, the trans-illumination light focused on the specimen
plane was adjusted to about $100 \, \mathrm{mW/cm^2}$ in all
the experiments. The sample was illuminated for at least
$15$ minutes before starting an experiment, so that the
BR was always in its light-adapted form \cite{lightBR}.

It has been shown
that the reverse phase evaporation technique used to incorporate
BR in large unilamellar vesicles (typically 
$200$ nm in diameter) results in an asymmetrical orientation
of the BR molecules across the lipid bilayer \cite{petitesves}. 
Consequently, for these vesicles,
a proton gradient builds up across the
lipid membrane upon activation, which inhibits BR pumping activity.
However, since the electroformation technique is symmetrical, we do not expect 
any asymmetry in the BR orientation, and thus we do not expect inhibition of 
the pumping activity. To be on the safe side, we have performed additional experiments
which were designed to cancel any proton gradient according to the following procedure. 
A classical way of suppressing the inhibitory proton
gradient, without incorporating any additional active molecule in 
the membrane, is to add KCl (potassium chloride)
to the solution, since protons can then codiffuse passively through
the bilayer in the form of HCl, and  since
chloride ions can diffuse inside the vesicle
to ensure electroneutrality.
In reference \cite{JB}, KCl was added to both the internal and external solutions
up to $2$ mM, a concentration above which electroformation of giant vesicles
fails, in order
to get rid of a possible proton gradient. Our results proved to be insensitive
to the addition of KCl .

\subsection{Micropipet experiments}

The micropipet technique developed
by Evans and coworkers
\cite{evans} allows a quantification of the excess surface area stored
in the membrane fluctuations by pulling it out with a micropipet
aspiration: when a pressure difference is applied, the membrane is 
put under tension and sucked inside the pipet.
The experimental set-up was built on an inverted microscope
equipped with a $40$x objective (N.A. 0.75 air Ph2 Plan
Neofluoar, Zeiss) for epi-fluorescence, phase contrast and differential
interference contrast (DIC) microscopy.
The transmission phase contrast or DIC images
were recorded by a CCD
(Charge Coupled Device) camera (Sony, Paris, France).
The sample cell was temperature controlled at $15 \, \mathrm{^oC}$ by a water
flow to limit evaporation of the solution (see Fig. \ref{schematic}, bottom).
\begin{figure}
\epsfxsize=7cm
\centerline{\epsfbox{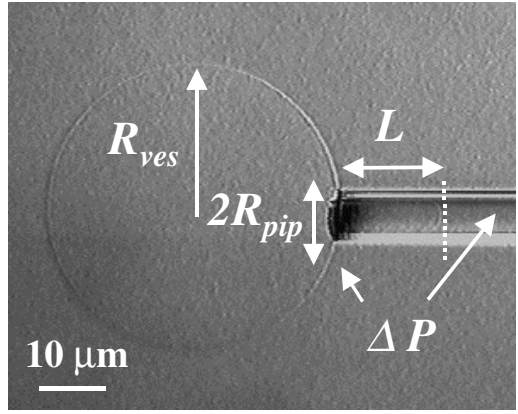}}
\caption[]{\label{micropip1}
Typical micropipet experiment.
The pressure difference $\Delta P$ is the difference between the
pressure outside and inside the pipet. The intrusion length $L$ is the length of membrane aspirated
inside the pipet when a pressure difference $\Delta P$ is applied.
The bar represents $10 \, \mathrm{\mu m}$.
}
\end{figure}
Glass micropipets were pulled from $1$ mm outer diameter borosilicate
capillaries (GC 100T-10; Phymep, Paris, France)
with a micropipet-puller (P-97; Sutter Instruments Co., Novato, California,
USA). The
micropipet tip was cut on a microforge
to obtain diameters up to $5-10 \, \mathrm{\mu m}$.
The pipets were treated with BSA (bovine serum albumin, $1 \, \%$) for
$30$ minutes to prevent adhesion of the lipid membrane to the glass
pipet walls. A pipet holder was mounted
on a three-dimensional piezo micromanipulator stage (Physik
Instrumente, Waldbron, Germany) in order to control the position of the pipet tip
within
$0.1 \, \mathrm{\mu m}$ accuracy.
The pressure difference $\Delta P$ between the outside and the inside
of the pipet was measured by a liquid-liquid pressure transducer
(DP103-20; Validyne, SEI3D, France)
with $0.01$ Pa accuracy. 
The pressure is imposed by a water height
difference between two water filled tanks equipped with micrometric displacements
(see Fig. \ref{schematic}, bottom). The absence of any air bubble in the water
circuit running from the tanks to the micropipet is crucial and
was checked before each experiment.
The relationship between the pressure difference $\Delta P$ and the imposed membrane tension $\sigma$
directly derives from Laplace's law \cite{evans}:
$$
\sigma = \frac{R_{pip}}{2(1-R_{pip}/R_{ves})} \Delta P
$$
where $R_{pip}$ is the pipet radius and $R_{ves}$ is the vesicle radius,
both measured directly from the DIC image
(see Fig. \ref{micropip1}).

The excess area stored in the membrane shape fluctuations $\alpha$ is
defined as  \mbox{$\alpha = (A-A_p)/A$,}
where $A$ is the actual area of the fluctuating membrane and $A_p$
is the area projected on the mean plane of the membrane.
During a micropipet experiment,
the excess area decreases as the membrane undulations are pulled out
by the increasing pressure difference. For a given $\Delta P$, an intrusion length
$L$ is aspirated inside the pipet (see Fig. \ref{micropip1}).
A reference state $(\Delta P_0, L_0)$ is defined as the lowest
suction pressure that can be applied in the experiment to aspirate the
fluctuating vesicle inside the micropipet \cite{evans}.
The variation of the excess area as compared to this reference state, the so-called
areal strain $\Delta \alpha = \alpha_0 - \alpha$, follows from
geometrical considerations. To first order in $\Delta L=L-L_0$, we have
\cite{evans}:
$$
\Delta \alpha = \alpha_0 - \alpha =
\frac{(R_{pip}/R_{ves})^2-(R_{pip}/R_{ves})^3}{2R_{pip}} \Delta L 
$$
The increase of the intrusion length $\Delta L$ was measured by
image analysis with pixel accuracy, {\it i.e.} $0.2 \,
\mathrm{\mu m}$ with the $40$x objective.

The excess area can be expressed using the local displacement
$u$ of the membrane around its mean plane \cite{ilnuovo}:
\begin{equation}
\alpha = \langle (\nabper u)^2 / 2 \rangle  = \frac{1}{(2 \pi)^2}
\int_{0}^{q_{max}} \frac{1}{2} \qpm^2
\langle u(\qp)^2 \rangle 2 \pi \qpm d\qpm
\label{alphadef}
\end{equation}
In the low q regime the convergence of the integral is guaranteed by a crossover from 
a curvature dominated regime $\spq = kT / \kappa \qpm^4$ to a tension dominated regime 
$\spq = kT / \sigma \qpm ^{2}$ for a passive membrane,
where $k$ is the Boltzmann constant, $T$ is the absolute temperature and
$\kappa$ is the bending modulus of the membrane. 
The upper limit is $q_{max}=2 \pi / a$,
where $a$ is a microscopic length.
In the
entropic regime, {\it i.e.} at low tension, inserting the
fluctuation spectrum of an equilibrium membrane 
gives the dependence of the excess area $\alpha$ as
a function of the membrane tension $\sigma$ \cite{evans,ilnuovo}:
$\alpha = (kT/8 \pi \kappa) \ln (cst/\sigma)$, where
$cst$ is an integration constant. The areal
strain $\Delta \alpha = \alpha_0 - \alpha$ is thus:
\begin{equation}
\Delta \alpha =\alpha_0 - \alpha =  \frac{kT}{8 \pi \kappa} \ln
\frac{\sigma}{\sigma_0} .
\label{alphapassive}
\end{equation}
For a passive membrane, the linear relationship between
the logarithm of the tension
$\ln \sigma$ and the areal strain $\Delta \alpha$ allows the
determination of the bending modulus $\kappa$.
We will give in section \ref{exptheor} a similar relation relevant to the active
case.

\subsection{Essential results}

\begin{figure}
\epsfysize=9cm
\centerline{\epsfbox{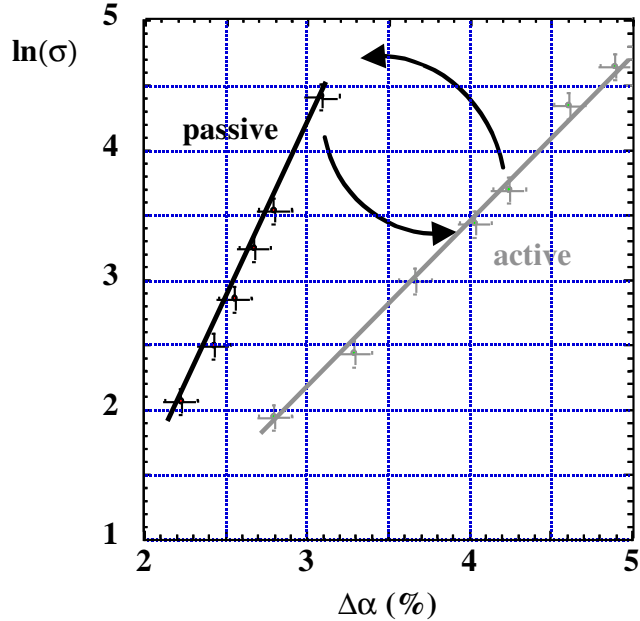}}
\caption[]{\label{exppip}
Variation of the logarithm of the tension $\sigma$ versus the areal 
strain $\Delta \alpha$ for the same vesicle containing BR, alternatively passive and 
active.
}
\end{figure}

\begin{figure}
\epsfxsize=7cm
\centerline{\epsfbox{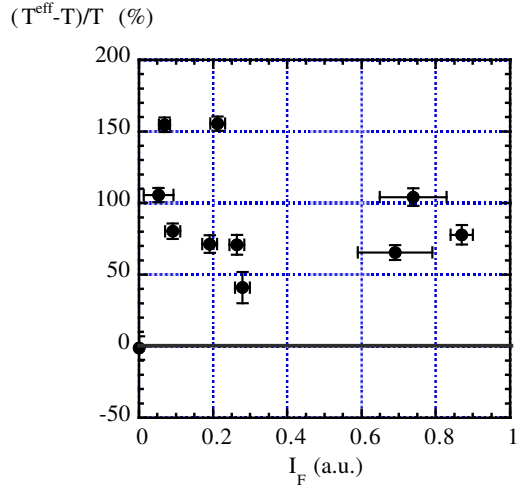}}
\caption[]{\label{Teffconc} 
Variation of the effective temperature $T^{eff}$ of the active membrane as a 
function of the fluorescence intensity $I_{F}$, thus of the BR 
concentration.
}
\end{figure}

The results reported in \cite{JB} and duplicated in Figure 
\ref{exppip}  
show that when the
vesicles are illuminated with green-yellow light, the
slope of the logarithm of tension versus the areal strain is smaller
than in the case where the vesicles are illuminated with red light.
This indicates that the excess area is larger when the BR is illuminated
with green-yellow light, and consequently that BR activity induces
an amplification of the membrane shape fluctuations. The quality of the fit
suggests that one can describe the effect of the BR activity in terms of 
an effective temperature $T^{eff} \simeq 2T$. 
An other important feature of the
experiment concerns the dependence of $T^{eff}$ on BR concentration. 
Fig. \ref{Teffconc} shows that in a concentration range 
that we estimate between approximatively $10^{15}$ and $10^{16} \, \mathrm{BR/m^{2}}$
the effective temperature is essentially 
independent on BR concentration. This may look surprising since the BR 
activity is the driving force.

Before developing the interpretation of these results, we must first guarantee 
their reliability, {\it i.e.} that it is an effect related to the out-of-equilibrium
pumping activity of BR and nothing else. The control experiments with pure lipid 
vesicles obviously exclude a role of the lipids themselves. For these, 
we find for both green-yellow and red illumination the expected $kT/\kappa \simeq 0.1$ value
\cite{EPC}.
Using simple estimates, we can also rule out the possibility of any
thermally induced artefact due
to the larger absorption of light by BR in the green-yellow wavelength.
Assuming that one BR molecule absorbs one photon each
$\tau \simeq 5$ ms,
the total stationary flux (total energy received per unit area of membrane
and per unit of time) is:
$$
W = \frac{h \nu}{\tau} \times \bar \rho
$$
where $h$ is the Planck constant, $\nu$ is the photon frequency
and $\bar \rho$ is the mean BR density. 
In a pessimistic estimate, we assume that this total flux $W$
is dissipated via conduction in the surrounding water. The sample cell
is temperature controlled by a cold water circulation and we assume
that a temperature gradient arises from the BR heating between the
membrane and the sample cell wall. This gradient extends over a
typical length $L=1$ mm which is the size of the sample cell.
With $C=4.18 \times 10^{6} \, \mathrm{J . m^{-3} . \, { ^o K}^{-1}}$ the heat 
capacity of water and $K_{T}=1.5 \times 10^{-7} \, \mathrm{m^2/s}$ 
the heat diffusion coefficient, we have:
$$
 C K_{T} \frac{\Delta T}{L} = W =  \frac{h \nu}{\tau} \times \bar 
 \rho
$$
and
$$
\Delta T = \frac{h \nu}{\tau}\times \bar \rho \times \frac{L}{C K_{T}}
$$
This yields a temperature increase of $ \Delta T = 2 \times 10^{-3} \,
\mathrm{^o K}$, five orders of magnitude smaller than the reported 
increase in effective temperature, which can not account for the observed effect
\cite{meleard97}.
Direct heating is just totally inefficient (note also that direct 
heating of water is clearly ruled out by the experiments on pure phospholipidic
vesicles).

Most importantly, the experiments with glycerol prove that
the observed effect is of nonequilibrium origin.
The addition of glycerol modifies dynamic parameters such as
the solvent viscosity $\eta$, the permeation coefficient
$\lambda_p$ and the active force $F_a$. At thermodynamic equilibrium,
such parameters cannot play a role in
the fluctuation spectrum 
as imposed by the fluctuation-dissipation theorem. For an active membrane
however, these parameters play a role as can be seen from 
references \cite{JP1} or \cite{JP3}.
The addition of glycerol increases the solvent viscosity while it decreases
its permeation coefficient. 
The results given in reference \cite{JB} report a lower increase
in the effective temperature when $16 \, \%$ and $25 \, \%$ 
(w/w) glycerol is added,
clearly revealing the out of equilibrium nature of the effect. 
This result is qualitatively consistent with the observation that the BR 
pumping activity is diminished upon addition of glycerol due to an increase in the 
lifetime of the M intermediate \cite{effetglyc}.
Finally, the fact that the observed effect does not depend on the
measuring sequence (red light experiment or green light experiment
first) rules out a potential role of the conformational change between
the light-adapted and the dark-adapted states \cite{lightBR}.
This is also consistent with the result that the ratio $kT/\kappa$ in the 
passive case is the same as that of the pure phospholipidic vesicles. The renormalisation of the bending
rigidity by the BR is not measurable. All these observations give strong support
to the assertion that the effect is indeed due to the proton pumping activity.

The use of an effective temperature to qualitatively
interpret the results according to equation (\ref{alphapassive})
is justified by
the good quality of the linear fits of the micropipet
experiments performed in \cite{JB}.
However, we need to develop a complete theoretical analysis to understand all 
these experimental features quantitatively.

\section{Theory}
\label{theor}

As in references \cite{JP1,JP2,JP3}, we consider situations in which a membrane under 
tension is subjected to random forces of two different origins. These arise   
(i) from thermal agitation, i.e., the Brownian motion of the membrane, simply
because membrane and solvent have a thermodynamic temperature,    
and (ii) from `biological' activity such as proton pumping
of the BR. The membrane equation of motion can be written
to lowest order:
\begin{equation}
\lambda_{P}^{-1}(\frac{\partial u}{\partial 
t}(\rp,t)-V_{z}(\rp,t))=\delta P(\rp,t) - \delta \Pi (\rp,t) +F_{a}\psi 
(\rp,t)+F_{a}^{\prime} \rho \Delta _{\perp } u(\rp,t) + f_{p} (\rp,t)    
\label{langevin}
\end{equation}
In this expression, $u(\rp,t)$ is the membrane displacement at point 
$\rp =(x,y)$ with respect 
to a $(x,y)$ reference plane orthogonal to ${\bf \hat{z}}$, the average membrane normal
and $\Delta_{\perp}$ is the Laplacian in the $xy$ plane. 
$V_{z}(\rp,t)$ is the fluid velocity, in the normal direction at the membrane surface;
$\lambda _{P}$ is the membrane permeation coefficient.  $\delta 
P(\rp,t)= P(\rp,z=0^{+},t)-P(\rp,z=0^{-},t)$ is the pressure difference 
across the membrane and $\delta 
\Pi(\rp,t)= \Pi (\rp,z=0^{+},t)- \Pi (\rp,z=0^{-},t)$ the osmotic 
pressure difference. This osmotic pressure difference results from 
the proton pumping activity: for each BR cycle 
one proton is transferred across the membrane. 
In principle the 
calculation of $\delta \Pi$ cannot be achieved  without solving all 
dynamical equations of the problem. However, considering the 
convective term of the proton flux as a second order correction allows 
to evaluate $\delta \Pi$ separately. We postpone this derivation to 
appendix \ref{app1}. $f_{p}$ is the brownian noise acting on the membrane
corresponding to the dissipation of energy in the permeation process
and satisfies:
$$
\langle f_{p}(\rp,t)\rangle = 0
$$
$$
\langle f_{p}(\rp,t) f_{p}(\rp^{\prime},t^{\prime})\rangle = 
2kT\lambda_{P}^{-1} \delta(\rp -{\bf {r_{\perp}^{\prime}}}) 
\delta(t - t^{\prime})  
$$
The term $F_{a} \psi (\rp,t) $ results from the BR activity. 
More precisely $\psi (\rp,t) = \rho ^{\uparrow} (\rp,t) - \rho ^{\downarrow} (\rp,t)$ 
is the local difference between the density $\rho ^{\uparrow} 
(\rp,t)$ of BR
molecules transferring protons in the  direction ${\bf \hat{z}}$ (up) and the 
density $\rho ^{\downarrow} (\rp,t)$ of BR molecules transfering protons 
in the -${\bf \hat{z}}$ direction (down). 
$F_{a}$ is the average elementary force transmitted to the membrane by a steady 
proton transfer. The flip-flop of BR is expected to be much slower than 
that of phospholipids and thus $\rho ^{\uparrow} (\rp,t)$ and
$\rho ^{\downarrow} (\rp,t)$ can be considered as separately conserved 
quantities. As explained in section \ref{expdetails}, with our experimental conditions, 
the probability of inserting a BR molecule into the phospholipid membrane 
does not depend on the pumping direction, thus $\langle \rho ^{\uparrow} 
\rangle = \langle \rho ^{\downarrow} \rangle $ and $\langle \psi 
\rangle=0$. $\lambda _{P} F_{a}$ can be understood as measuring the average volume 
transferred through the membrane per BR and per unit time. 
Thus, $\lambda_{P} F_{a} \psi (\rp,t)$, 
which we will refer to as the `active-permeative' term, 
can also be understood as the local volume 
transferred through the membrane per unit area and per unit time due to the pumping 
imbalance between up and down BR molecules. Terms corresponding to the stochastic 
nature of the pumping activity have been omitted since they have been shown 
to lead to
smaller effects than those due to the collective $\psi$ fluctuations
\cite{JP1,JP2}. 
As discussed in \cite{JP3}, the fourth term describes the fact that pumping may work 
better when the membrane is curved with a given sign: $F_{a}^{\prime}$ 
measures this sensitivity per pump and $\rho = \rho ^{\uparrow} (\rp,t) + 
\rho ^{\downarrow} (\rp,t)$. 
There is experimental evidence for such a coupling in the functioning of
certain ion-channels called TRAAK and TREK \cite{trak}. 

We further need equations for the fluid and for the BR density dynamics. 
Navier-Stokes equations have to be implemented in two ways: one first has to keep track 
of the Laplace force exerted by the membrane on the fluid,  and this can
be done in the usual way \cite{JP2}. 

Secondly, each BR has a small but finite spatial extent. Its activity will disturb the
ambient solvent in the form of a distribution of force densities in its vicinity. 
Since no external force source is present, 
the total force must vanish, but its first moment will in general be 
present. For convenience, we adopt
the simplest set of force-densities consistent with this requirement: a pair of 
oppositely directed point forces, separated by a distance of order the 
size of a BR molecule. This implies a dipolar force density  
$F_{a} \left[ \delta (z - w^{\uparrow })-\delta (z+w^{\downarrow }) \right] 
\psi (\rp,t)$ 
in the Stokes equations (this term will be called the `active-hydrodynamic' term). 
$w^{\uparrow }$ and $w^{\downarrow }$ are 
lengths of the order of the BR size; their values are {\it a priori} 
unequal since the BR, or any molecule with unidirectional activity, is not up-down
symmetric. Similarly, the term \( F_{a}^{\prime} \left[ \delta 
(z-w^{\uparrow}) - \delta (z+w^{\downarrow}\right] \rho \Delta_{\perp} u(\rp,t) \) 
describing the sensitivity of the force dipole to curvature should be 
kept. Thus, in the low Reynolds number regime appropriate to these experiments, 
we can write:
\begin{equation}
\begin{array}{rcl}
0 & = & - {\bf \nabla} P({\bf r}, t) - \frac {\delta F} {\delta u} (\rp,t) \delta (z) {\bf \hat{z}} \\ [6pt]
  & & \quad + F_{a} \left[ \delta (z-w^{\uparrow}) - \delta (z+w^{\downarrow}) 
  \right] \psi ({\bf r}, t)  {\bf \hat{z}} \\ [6pt]
  & &  \qquad + F_{a}^{\prime} \Delta_{\perp}´ u \left[ \delta 
(z-w^{\uparrow}) - \delta (z+w^{\downarrow}) \right] \rho (\rp,t) {\bf \hat{z}} \\ [6pt] 
  & & \qquad + \eta \Delta {\bf V} ({\bf r}, t) + {\bf f}_{h} ({\bf r}, t),  
\end{array}
\label{navier}
\end{equation}
where ${\bf r}$ refers to the three-dimensional position vector, and $\rp$ has the same 
meaning as in
(\ref{langevin}). $P({\bf r},t)$ is the three-dimensional pressure field, ${\bf V} ({\bf r},t)$ the 
three-dimensional fluid velocity field, $F$ is the membrane free energy: 
$$
F = \frac{1}{2} \int d^{2} \rp 
\left[ \kappa (\Delta_{\perp} u(\rp ))^2  + \sigma  ({\bf \nabla} u(\rp ))^2 - 
 2 \Xi \psi (\rp ) \Delta_{\perp} u + \chi \psi^{2} (\rp ) \right] \ 
$$
$\kappa$ is the membrane `bare' bending modulus, $\sigma$ the membrane 
tension, $\Xi$ a coefficient linking membrane curvature and BR
imbalance, $\chi$ the `bare' susceptibility corresponding to that 
imbalance 
(for small enough densities $\chi \approx kT / \rho$). Orders of magnitude will be 
given in the next section. The third term of equation (\ref{navier}), 
is the `force dipole' density already described, 
the fifth and sixth are the usual viscous terms and associated forces:
\[
\begin{array}{c}
\langle {\bf f}_{h} ({\bf r},t) \rangle = 0 \\[6pt]
\langle {f_{hi} ({\bf r},t)} {f_{hj} 
({\bf r^{\prime}},t^{\prime})} \rangle = 2 kT \eta \: \{ -\delta 
_{ij} \nabla ^{2} + \partial _{i} \partial _{j} \} \delta ({\bf r}-{\bf r^\prime })
 \delta (t-t^\prime )
\end{array}
\]

Last, we need a dynamical equation for the BR imbalance 
density $\psi$. Following \cite{JP3}, we can write in the linear regime:
\begin{equation}
\frac {\partial \psi} {\partial t} = \Lambda \Delta_{\perp} \frac 
{\delta F} {\delta \psi} + \nabper \cdot {\bf f}_{\psi} 
\label {unbalance}
\end{equation}
with $\Lambda = D / \chi$, where $D$ is the diffusion coefficient of the BR
molecules in the membrane.
This expression is valid for $\langle \psi \rangle =0$, in the absence of 
fluctuation corrections, 
and the last term of (\ref{unbalance}) is a conserving noise, i.e., the divergence of a
random current with  
\[
\begin{array}{c}
\langle {\bf f}_{\psi }({\bf r_{\perp }},t) \rangle =0  \\ [6pt]
\langle f_{\psi i } ({\bf r_{\perp }},t) f_{\psi j } ({\bf r_{\perp }} 
^{\prime},t^{\prime}) \rangle = 2 \Lambda kT \delta_{ij} 
\: \delta ({\bf r_{\perp }}-{\bf r_{\perp } 
^{\prime}}) \delta (t-t^\prime) 
\end{array}
\]

In order to compare experiment and theory, we need to calculate the equal time 
correlation function $\langle u(\qp ,t) u(\qp^\prime ,t) 
\rangle $. We first eliminate $V_z$ in (\ref{langevin}) by solving for it from the 
Stokes equation (\ref{navier}) in Fourier space to get:
\begin{equation}
\frac {\partial u} {\partial t} (\qp ,t)+ \tau _{u} ^{-1} u(\qp 
,t)=\beta \: \psi (\qp ,t) + \mu   
\label{fournavier1}
\end{equation}

\begin{equation}
\frac {\partial \psi} {\partial t} (\qp ,t)+ \tau _{\psi } ^{-1} \psi (\qp 
,t)=\gamma \: u(\qp ,t) + \nu  
\label{fournavier2}
\end{equation}
in which we have chosen the convention 
\[
f(\qp ,t)= \int f(\rp ,t) \: \exp(i\qp \cdot \rp) \,  d^{2} \rp 
\]
for Fourier transforms, 
and used \( \delta \Pi \simeq 2 kT \dot{a} \psi (\qp ,t) / D_{P} \qpm 
\)
where $\dot{a}$ is the proton pumping rate and $D_{P}$ an effective 
proton diffusion coefficient (see appendix \ref{app1}). 
The parameters entering equations (\ref{fournavier1}) and ({\ref{fournavier2})
are listed below:
\[
\begin{array}{lr}
\tau _{u} ^{-1} = (\lambda _{P} + \frac {1} {4 \eta \qpm}) (\sigma 
\qpm ^{2} + \kappa \qpm ^{4}) + \rho \lambda _{P} F_{a}^{\prime } \qpm 
^{2} - \frac {\rho {\mathcal{P}}_{a}^{\prime} w} {4 \eta } \qpm ^{3} & \\ [6pt]
\tau _{\psi } ^{-1} = \Lambda \chi \qpm ^{2} = D \qpm ^{2} & \qquad 
\bar{F_{a}} = F_{a} - 2kT \dot{a} / D_{P} \qpm  \\ [6pt]
\beta = \lambda _{P} \bar{F_{a}} - {\mathcal{P}}_{a} \frac {\qpm w} {4 
\eta } - \Xi \qpm ^{2} (\lambda _{P} + \frac {1} {4 \eta \qpm})  & \qquad 
{\mathcal{P}}_{a} = F_{a} \frac {w_{\uparrow}^{2} - 
w_{\downarrow}^{2}} {2w} \\ [6pt] 
\mu = \lambda _{P} f_{p} (\qp ,t) + \frac {1} {2\pi \eta}  \int  \frac 
{ {\bf f}_{h}({\bf q} ,t) \cdot {\bf \hat{z}} } {q^{2}}  dq_{z}  & \qquad {\mathcal{P}}_{a} ^{\prime} 
= F_{a}^{\prime} \frac {w_{\uparrow}^{2} - w_{\downarrow}^{2}} {2w} \\
\gamma = - \Lambda \Xi \qpm ^{4} & \\
\nu = -i \qp \: {\bf f}_{\psi} (\qp ,t) & 
\end{array}
\]
Calculating the equal time $\langle u(\qp ,t) u(\qp^\prime ,t) 
\rangle $ correlation function is 
straightforward although somewhat tedious; we find 
\begin{equation}
\langle u (\qp ,t) u(-\qp ,t) \rangle = \frac {kT} {(\tau _{u}^{-1}+\tau 
_{\psi }^{-1}) ({\tau _{u}^{e}}^{-1} + \tau _{a}^{-1})} \left[ \frac 
{(\lambda_{P} F_{a}-{\mathcal{P}}_{a} \frac {\qpm w} {4 \eta})^{2}} 
{\chi} + (\tau _{u}^{-1}+\tau 
_{\psi }^{-1} + \tau _{a}^{-1}) (\lambda _{P} +\frac {1} {4 \eta 
\qpm }) \right]
\label{correlation}
\end{equation}
with 
$$
\tau _{a}^{-1}= \frac {\Xi } {\chi } \: \qpm ^{2} \: \left ( \lambda_{P} F_{a} -{\mathcal{P}}_{a} 
\frac {\qpm w} {4 \eta}\right )
$$
${\tau _{u} ^{e}}^{-1}$ has exactly the same structure than $\tau _{u} ^{-1}$, but 
$\kappa$ is replaced by \( \kappa ^{e} = \kappa - \Xi ^{2} / \chi \).
Note that in the absence of active noise, (\ref{correlation}) reduces to its thermal
equilibrium expression, as it should. \\

In the long wavelength limit, one finds for a membrane under tension 
(neglecting the osmotic contribution):
\begin{equation}
\langle u (\qp ,t) u(-\qp ,t) \rangle = \frac {k T^{e}} {\sigma \qpm^{2}}
\label{tension}
\end{equation}
with \( T^{e}=T \; (1+16 \lambda _{P}^{2} F_{a}^{2} \eta ^{2} / \chi 
\sigma ) \) 
which means that a tense membrane is flat at longwavelength, 
even in the presence of non-thermal noise, and that one can define
an effective temperature higher than the actual one. \\

For a tensionless membrane, we now find for $\qpm \rightarrow 0$:
\begin{equation}
\langle u (\qp ,t) u(-\qp ,t) \rangle = \frac {k {T^{\prime}}^{e}} 
{\kappa \qpm^{4}}
\label{notension}
\end{equation}
with \( {T^{\prime}}^{e} = T \cdot (\lambda _{P} F_{a})^2 \kappa / [ \chi (D+\rho 
\lambda _{P} F_{a}^{\prime} ) ( \rho \lambda _{P} F_{a}^{\prime} + 
\frac {\Xi} {\chi} \lambda _{P} F_{a} ) ] \).
This expression is equivalent to the one given in \cite{JP3}, 
and in particular the effective temperature does not depend 
on the pumping density for small enough densities. At long enough 
wavelength, the osmotic term should always dominate, but we show 
in the next section that the experimentally relevant regimes in fact 
imply the contribution of the force dipoles.

\section{Experimentally relevant regime and discussion}
\label{exptheor}

Let us first point out that in all equations the active-permeative and 
the active-hydrodynamic terms come in the combination:
\[
\lambda _{P} F_{a} - \frac {{\mathcal{P}}_{a}} {4 \eta} {\qpm w}
\]
This tells us that in the long wavelength limit, the active-permeative  
term always dominates over the  active-hydrodynamic term. However, the crossover
wavevector, below which the active-permeative term wins reads:
\[
q_{\perp c}= \frac {4 (\eta \lambda_{P}) F_{a}} {w {\mathcal{P}}_{a}}
\]
or for the corresponding length:
\[
l_{\perp c}= 2 \pi \frac {w} {4 \eta \lambda_{P}} \cdot (\frac {{\mathcal{P}}_{a}}{F_{a}})
\]
That is with $\eta \lambda_{P} = l_{P}$ the permeation length, and 
\( {\mathcal{P}}_{a}/{F_{a}} \simeq w \):
$$
l_{\perp c} \simeq 2 \pi \frac {w^{2}} {4l_{P}}
$$
At first sight, one might be tempted to state that this crossover 
length is microscopic, but it turns out that $l_{P}$ is of the order 
of, or smaller than, a Fermi. Indeed, with $\eta = 10^{-3}$ kg/m.s
and \( \lambda _{P} \lesssim 10^{-12} \, \mathrm{m^3/N.s} \) \cite{permeation} , we 
find $l_{P} \lesssim 10^{-15}$ m. 
Then, with $w \simeq 5 \cdot 10^{-9}$ m, we find:
$$
l_{\perp c} \simeq 3 \cdot 10^{-2} \, \mathrm{m}
$$
As a result, all active-permeation terms may be omitted in the micron and 
submicron length scales we are dealing with
($ 2 \pi / R_{ves} \simeq 3 \cdot 10^{5} \,  {\mathrm m}^-1 \lesssim q \lesssim 2 \pi /a 
\simeq 6 \cdot 10^{10} \, {\mathrm m}^-1$, with $R_{ves} \simeq 20 \, \mathrm{\mu m}$
and $a \simeq 0.1 \,$ nm).

The osmotic contribution 
is negligible as well. To see this, compare $2 \lambda _{P} 
kT \dot{a}/D_{P}\qpm $ and ${\mathcal{P}}_{a} \qpm w/4 \eta $: this yields the 
cross over wavevector 
$$
q_{\perp c}^{\prime} = \left( \frac {8 \lambda _{P} \eta kT \dot{a}} 
{{\mathcal{P}}_{a} w D_{P}} \right) ^{1/2}
$$
or the cross over length
$$
l_{\perp c}^{\prime} = 2 \pi \left( \frac {{\mathcal{P}}_{a} w D_{P}} {8 \lambda _{P} \eta kT \dot{a}} 
 \right) ^{1/2}
$$
With an effective proton diffusivity $D_{P} 
\simeq 10^{-9} \, \mathrm{m^{2}/s}$, a pumping rate $\dot{a} = 
10^{3} \, \mathrm{s^{-1}}$, ${\mathcal P}_{a} \simeq \kappa \simeq 10 kT$  
(see appendix \ref{app2}),
and other material parameters as above,  
one finds $l_{\perp c}^{\prime} \simeq 10^{-2}$ m:
the osmotic contribution is totally negligible as well. In 
the case of ion channels for which  $\dot{a} = 
10^{7} \, \mathrm{s^{-1}}$, the cross over length is reduced by a factor 
hundred, that is to a few tens of microns; this may be accessible to 
experiment. 
Similarly, since $\l_P \ll q^{-1}$, terms arising 
from permeative friction may entirely be omitted in the equations. 
We have now:
\[
\begin{array}{lr}
\tau_{u}^{-1} = \frac {1} {4 \eta} (\sigma \qpm + \tilde{\kappa} 
\qpm^{3}) & \qquad \text{with} \: \tilde{\kappa}= \kappa - \rho 
{\mathcal{P}}_{a}^{\prime} w \\ [6pt]
{\tau_{u}^{e}}^{-1}  = \frac {1} {4 \eta} (\sigma \qpm + 
\tilde{\kappa ^{e}} 
\qpm^{3}) & \qquad \quad \tilde{\kappa ^{e}}= \kappa^{e} - \rho 
{\mathcal{P}}_{a}^{\prime} w \\ [6pt]
\tau_{\psi }^{-1} = D \qpm ^{2} & \\ [6pt]
\tau_{a}^{-1} = - \frac {\Xi} {4 \chi \eta} {\mathcal{P}}_{a} w \qpm^{3} & \\
\end{array}
\]
A further simplification can be obtained with the remark that in our 
experimental conditions $D \ll {(\sigma \kappa )}^{1/2} \eta^{-1} $ ({\it 
i.e.},
more precisely, \( D \lesssim 10^{-2} {(\sigma \kappa )}^{1/2} \eta^{-1} 
\) with $D \simeq 10^{-12} \, \mathrm{m^2/s}$ \cite{diffusion}).
This means that one can further ignore the diffusion term in the 
$\langle u (\qp ,t) u(-\qp ,t) \rangle$ correlation function. Under such conditions, 
equation (\ref{correlation}) reduces to the expression:
\begin{equation}
\langle u (\qp ,t) u(-\qp ,t) \rangle \simeq \frac {kT} {\sigma \qpm 
^{2} + \tilde{\tilde{\kappa ^{e}}} \qpm ^{4}} + \frac {kT \left[ 
{\mathcal{P}}_{a}^{2} w^{2} - \Xi {\mathcal{P}}_{a} w\right]} {\chi 
(\sigma + \tilde{\kappa} \qpm^{2}) (\sigma + \tilde{\tilde{\kappa 
^{e}}} \qpm^{2})}
\label{reducecor}
\end{equation}
where \( \tilde{\tilde{\kappa ^{e}}} = \tilde{\kappa ^{e}} - 
{\mathcal{P}}_{a} w \Xi / \chi \) \\
From this correlation function, we can calculate the relationship between the areal strain and 
the membrane tension: 
\begin{equation}
\Delta \alpha = \alpha_{0} - \alpha = \frac{kT}{8 \pi} \left( \frac 
{{\mathcal{P}}_{a}^{2} w^{2} - \Xi {\mathcal{P}}_{a} w} {\tilde{\tilde{\kappa 
^{e}}} \tilde{\kappa} \chi} + \frac{1}{\tilde{\tilde{\kappa 
^{e}}}} \right) \ln \left( \frac {\sigma} {\sigma_{0}} \right)
\label{actalpha}
\end{equation}
or
\begin{equation}
\Delta \alpha = \alpha_{0} - \alpha = \frac {kT^{eff}} {8 \pi \kappa}
\ln (\frac {\sigma} {\sigma_{0}} )
\label{actalpharedu}
\end{equation}
with
\begin{equation}
\frac {T^{eff}} {T} = \frac {\kappa} {\tilde{\tilde{\kappa ^{e}}}}
\left( 1+ \frac {{\mathcal{P}}_{a}^{2} w^{2} - \Xi {\mathcal{P}}_{a} w} 
{\tilde{\kappa} \chi} \right)
\label{Teff} 
\end{equation}
The functional relation (\ref{actalpharedu}) is identical to the one 
holding in the equilibrium case (except that the `temperature' is an effective
nonequilibrium noise level) and is clearly compatible with the
experimental results. The whole theory accounts well for the experimental observations  
if we note in addition the following four results:

\begin{itemize}
\item the value of the bending modulus, measured in the passive case, is 
insensitive to the BR concentration  and equal 
to that of pure phospholipidic membrane,  
\item the effective temperature is about twice as large as the actual 
temperature, 
\item the effective temperature is essentially independent of the 
BR surface concentration in the investigated domain,
\item the reduced effective temperature difference $(T^{eff}-T)/T$ 
decreases by a factor 3 when 25 $\%$  glycerol is  added. 
\end{itemize}
The first observation is easily explained, as detailed in appendix 
\ref{app2}. In order to discuss the second and third observations one 
must estimate the different terms entering equation (\ref{Teff}). We 
provide details on these estimates in appendix \ref{app2}. We 
expect:

\[ 
\begin{array}{c}
| {\mathcal{P}}_{a} | \simeq \kappa \\ [6 pt]
| \Xi | \simeq wkT \\ [6 pt]
| {\mathcal{P}}_{a}^{\prime} | \simeq \kappa w \\ [6 pt]
\chi \simeq kT/ \rho \simeq kTl^{2} 
\end{array}
\]
where $l$ is the average distance between BR
molecules. We have been able to vary the concentration $\rho$ over 
approximately one order of magnitude, that is roughly $l$ from $w$ to 
$3w$. With such estimates, and chosing the signs in such a way that 
the system is stable, we expect:
$$
\frac {T^{eff}} {T} \simeq \frac {1+ \left( \frac {\kappa} {kT} 
+2 \right) \frac {w^{2}} {l^{2}} } {\left( 1+ \frac {w^{2}} {l^{2}} 
\right) \left( 1+ \left( 2- \frac {kT} {\kappa} \right) \frac 
{w^{2}} {l^{2}} \right)}
$$
with $0.3 \leq w / l \leq 1 $ and knowing that $\kappa \simeq 
10kT$, we find:
$$
1.7 \lesssim \frac {T^{eff}} {T} \lesssim 2.3
$$
which is in very reasonable agreement with experiment. Of course, the 
numbers chosen above have some degree of arbitrariness, but 
one can change them appreciably while retaining a  
ratio $T^{eff} / T$ of order 2. For instance, $\Xi $ may be set 
to zero, keeping other values unchanged, and one finds 
$2 < T^{eff} / T < 2.7 $, which is less satisfactory but not 
off scale. 

Let us now turn to the glycerol dependence. It 
has been measured that a 25 $\%$  glycerol addition to water reduces 
the pumping activity of the bacteriorhodopsin by a factor $2.5$ 
\cite{effetglyc}. It is thus clear that both ${\mathcal{P}}_{a}$ and 
${\mathcal{P}}_{a}^{\prime}$ (and perhaps $\Xi$) have to be reduced by 
a factor $2.5$; all other parameters are essentially unchanged, as 
shown by the experiments performed with red light. The same type of 
estimate as before give the expected reduction of $(T^{eff}-T)/T$ by a 
factor three. \\

The net conclusion is thus that our analysis provides a satisfactory 
account of the experiment, although it is not able to pinpoint 
accurately values for $\Xi $, ${\mathcal{P}}_{a}^{\prime}$ and 
${\mathcal{P}}_{a}$. A more accurate experiment should reveal that the 
effective temperature should depend on BR density in a non 
trivial way. At low density, the effective temperature should be 
essentially equal to the actual temperature; it should increase 
proportionally to the density at moderate densities; eventually, at 
larger densities, it could even decrease after going through a 
maximum.
Together with an independent measurement of 
$\chi $, it should allow us to measure at least ${\mathcal{P}}_{a}$ and 
${\mathcal{P}}_{a}^{\prime}$. Our current accuracy does not allow for 
such a detailed analysis. \\

Thus, the proposed analysis gives a natural interpretation of the 
experimental data. It is one of those intriguing cases in which 
terms nominally subdominant in wavenumber provide by far 
the leading contribution. This pecularity is due to the very small 
value of the permeation coefficient: in the experimentally accessible 
domain, the membrane is practically impermeable and the effects due 
directly to the force exerted by the active centers on the membrane are negligible.

\appendix
\section{Osmotic pressure difference}
\label{app1}

Since BR selectively pumps protons, we just have to 
consider the osmotic pressure resulting from the three dimensional 
proton density $n({\bf r},t)$:
\begin{equation} 
\Pi = kT n({\bf r},t)
\label{osmotic}
\end{equation}
The protons dynamics is described as usual by conservation equations 
in each half space, above and below the membrane:
\begin{equation} 
\frac {\partial n} {\partial t} ({\bf r},t) + {\bf \nabla} \cdot {\bf J}_{n} =0
\label{conservation}
\end{equation}
$$
{\bf J}_{n} = n {\bf V} - D_{P} {\bf \nabla} n
$$
At the membrane, the coarsegrained proton flux ${\bf J}_{n} \cdot 
{\bf \hat{z}}$ in the membrane normal direction ${\bf \hat{z}}$, is 
given by the BR active transport:
$$
{\bf J}_{n} \cdot {\bf \hat{z}} = \psi ({\bf r},t) \: \dot{a}
$$
where $\dot{a}$ is the pumping rate, and $\psi ({\bf r},t)$ is 
defined in the main text. For a macroscopically symmetric membrane, 
$\psi$, $n$ and ${\bf V}$ are `small' fluctuating quantities. So the 
convective term can be omitted, as a second order correction. Now to 
the linear order, equation (\ref{conservation}) becomes in `hybrid' Fourier space (with 
obvious notations):
\begin{equation}
\begin{array}{c}
i \omega n(z,\qp ,\omega) + D_{P} \qpm^{2} n(z,\qp ,\omega) - 
D_{P} \frac{\partial^{2}} {{\partial z}^{2}} n(z,\qp ,\omega) =0  \\ [6pt]
-D_{P} \frac {\partial n} {\partial z} (z=0^{+}, \qp ,\omega) = 
-D_{P} \frac {\partial n} {\partial z} (z=0^{-},\qp ,\omega) = \psi 
(\qp,\omega) \dot{a}
\end{array}
\end{equation}
The solution to this problem is straightforward. One finds:
$$
n(z=0^{+},\qp ,\omega) - n(z=0^{-}, \qp, \omega)= \frac {2 \psi (\qp 
,\omega ) \dot{a}} {D_{P} (i \frac {\omega} {D_{P}} + \qpm ^{2})^{1/2}}
$$
so that 
\begin{equation}
\delta \Pi (\qp ,\omega) = \frac {2 kT \psi (\qp ,\omega) \dot{a}} 
{D_{P} {(\frac {i \omega} {D_{P}} + \qpm ^{2})}^{1/2}}
\label{pressosmot}
\end{equation}
The typical frequency over which $\psi (\qp ,\omega)$ varies is $D 
\qpm^{2}$. Eventhough $D_{P}$ is an effective diffusion coefficient 
renormalized by the time the proton spends attached to the 
hydrazoic acid resulting from the conversion of sodium azide in 
solution \cite{azide}, 
one always has $D_{P} \gg D$, and thus the term $\omega / D_{P}$ 
may be safely omitted in equation (\ref{pressosmot}). Then, one can 
equivalently write: 
\begin{equation}
\delta \Pi (\qp ,\omega) \simeq \frac {2 kT \psi (\qp ,\omega) \dot{a}} 
{D_{P} \qpm}
\end{equation}

\section{Orders of magnitude of the theoretical parameters}
\label{app2}

The long wavelength effective membrane curvature modulus $\kappa 
^{e}$, is as shown in section \ref{theor}:
\begin{equation}
\kappa^{e}= \kappa - \frac {\Xi ^{2}} {\chi}
\end{equation}
It is easy to convince oneself that the coupling term $\Xi$ in fact 
depends on the pumping activity. Let us first consider the passive 
case, and call the corresponding coeffiecient $\Xi _{p}$. A 
BR molecule with a given orientation may 'prefer' a given 
curvature sign for several different reasons. The first and most 
obvious  one is linked to a putative wedge shape. In such a case, one 
expects:
$$
\Xi_{p} \simeq \kappa R \theta
$$
in which $R$ is the 'radius' of the BR molecule and $\theta$ the 
wedge angle. The experiments performed with red light tell us:
$$
\frac {\Xi_{p}^{2}} {\chi} \ll \kappa
$$
That is: 
$$
\theta ^{2} \ll \frac {kT} {\kappa} \left( {\frac{l}{R}}\right) ^{2} 
$$
At the highest densities $l \simeq w \simeq R$, so with  $\kappa = 10 
kT$ the experiment requires $\theta \ll 1/3$, which is obviously a 
`weak' requirement: the absence of up-down symmetry in 
BR requires the existence of a wedge, but inspection 
of the molecular structure suggests that it is very small (for 
instance, it is very hard to coin a sign to it). So, the `steric' 
contribution to $\Xi_{p}$ can be safely neglected.

There can however, be other contributions and the next most obvious 
one results from flexoelectricity: a curved membrane generates an 
electric polarisation, hence a transmembrane electric field. Again, 
the absence of up-down symmetry in BR tells us that it 
must have a non zero electric dipole. The energy of the dipole in 
this transmembrane field, provides the coupling between curvature 
and $\psi$.
With the usual definitions, the transmembrane electric field can be 
written:
\begin{equation}
E = - \frac {e} {\epsilon w} \Delta_{\perp} u
\end{equation}
in which $\epsilon$ is the dielectric permittivity of the hydrophobic layer and
$e$ is the flexoelectric coefficient discussed by Petrov for 
instance \cite{petrov}. 
If we call $p$ the 
BR average longitudinal dipole, one has:
$$
\Xi _{p}^{f} = \frac {e p} {\epsilon w}
$$
The flexoelectric coefficient $e$ is a measured quantity \cite{petrov}:
$$
|e| \simeq 1.3 \cdot 10^{-20} \; \mathrm{C}
$$
Estimating $p \simeq \text{a few} \, q \cdot \delta $, in 
which $q$ is a unit charge and $\delta$ a distance of the order of a 
fraction of the membrane thickness, {\it{e.g.}} the hydrophilic part 
(note that it cannot be much larger otherwise the BR
would not be membrane soluble), 
and taking the dielectric permittivity of the hydrophobic layer
of the order of $\epsilon \simeq 3 \, \epsilon_0$, with 
the dielectric permittivity of vacuum $\epsilon_0 \simeq 8.8 \cdot
10^{-12} \, \mathrm{F . m^{-1}}$, we find:
$$
\frac {\Xi_{p}^{f \, 2}} {\kappa \chi} < 10^{-2}
$$
So in this case as well, one does find $\kappa ^{eff} \simeq \kappa$.

When the BR undergoes its pumping activity, the 
flexoelectric energy is dominated by the time average energy of the 
proton in the flexoelectric potential $e \Delta _{\perp} u / \epsilon$; 
assuming a duty ratio of one tenth, we expect then:
$$
\Xi_{a}^{f} \simeq \frac {q e} {10 \cdot \epsilon } \simeq \frac {1} 
{100 \cdot \epsilon} q^{2} \simeq w \; kT
$$

The force dipole ${\mathcal{P}}_{a}$ has the dimensions of an energy, 
and hence must be a fraction of the green-yellow photon energy. As a rough 
rule of thumb, we take again a duty ratio of a tenth:
$$
{\mathcal{P}}_{a} \simeq \frac {h \nu} {10} \simeq \kappa
$$
The curvature dependence of the force dipole can be estimated in a way 
similar to that used for $\Xi $. During its pumping cycle, the 
BR/proton system has probably to overcome a potential 
barrier $W_{b}$. The pumping rate is then controled by a Boltzmann 
factor $\exp(-W_{b}/kT)$. In the presence of curvature, the barrier is 
modified by the energy 
of the proton in the flexoelectric potential at the barrier location. 
We call $xW_{b}$ this location. Thus, the activity is multiplied by a 
factor $\exp \left[ -x ( e q \Delta _{\perp} u) / (\epsilon \cdot kT) 
\right]$ which can be linearized for small curvatures. Hence, we 
expect (with $x$ of the order of a few tenths):
$$
{\mathcal{P}}_{a}^{\prime} \simeq - {\mathcal{P}}_{a} x \frac {eq} 
{\epsilon \cdot kT} \simeq - \, \text{a few} \, w \cdot \kappa 
$$
Note that one could in principle estimate this coefficient by 
measuring the pumping activity in liposomes, as a function of 
liposome radius: the net result would strongly depend on the value of 
$x$. For $x \simeq$ a few tenths, one would need \% accuracy to 
measure the curvature dependence. For $x \simeq 1$, the effects would 
be much larger and the exponential nature of the relation should start 
to show up. However, even if the direct  effect is not easily 
measurable, the incidence on formula (\ref{Teff}) can be important.




\begin{references}

\bibitem{ReviewsMbBio}
See for instance, {\it Molecular Biology of the Cell}, Ch. 10
(Third Ed.) by B. Alberts {\it et al.} (Garland Publishing,
New York, 1994).

\bibitem{ReviewsMbPhy}
{\it Structure and Dynamics of
Membranes}, edited by R. Lipowsky, E. Sackmann (Elsevier North
Holland, Amsterdam, 1995); M. Bloom, E. Evans, and O. G. Mouritsen,
Q. Rev. Biophys. {\bf 24}, 293--297 (1991) and references therein;
R. Lipowsky, Nature {\bf 349}, 475--481 (1991) and references therein.

\bibitem{JB} J-B. Manneville, P. Bassereau, D. L\'evy, and J. Prost,
Phys. Rev. Lett. {\bf 82}, 4356--4359 (1999).

\bibitem{JP1} J. Prost and R. Bruinsma,
Europhys. Lett. {\bf 33}, 321--326 (1996).

\bibitem{JP2} J. Prost, J-B. Manneville, and R. Bruinsma,
Eur. Phys. J. B {\bf 1}, 465--481 (1998).

\bibitem{JP3} S. Ramaswamy, J. Toner, and J. Prost,
Pramana-J. Phys. {\bf 53}, 237--242 (1999);
S. Ramaswamy, J. Toner, and J. Prost,
Phys. Rev. Lett. {\bf 84}, 3494--3497 (2000). 



\bibitem{ReviewsBR1} R. Henderson and P. N. T. Unwin,
Nature {\bf 257}, 28--32 (1975).

\bibitem{ReviewsBR2} D. Oesterhelt and W. Stoeckenius,
Methods Enzymol. {\bf 31}, 667--678 (1974).

\bibitem{structBR} Y. Kimura {\it et al.},
Nature {\bf 389}, 206--211 (1997) ; E. Pebay-Peyroula
{\it et al.}, Science {\bf 277}, 1676--1681 (1997).

\bibitem{ReviewsBR3} For recent reviews see:
U. Haupts, J. Tittor and D. Oesterhelt, 
Ann. Rev. Biophys. Biomol. Struct. {\bf 28}, 367--399 (1999);
S. Subramaniam {\it et al.}, J. Mol. Biol. {\bf 287}, 145--161 (1999);
J. K. Lanyi, J. Biol. Chem. {\bf 272}, 31209--31212 (1997); 
W. Kuhlbrandt, Nature  {\bf 406} 569--570 (2000); J. Heberle {\it et al.}, 
Biophys. Chem. {\bf 85} 229--248 (2000); N.A. Dencher {\it et al.} 
Biochim. Biophys. Acta {\bf 1460} 192--203 (2000).

\bibitem{pumpBR} See for example:
H. Luecke {\it et al.},
Science {\bf 286}, 255--260 (1999);
H. Luecke, H.-T. Richter, and J. K. Lanyi,
Science {\bf 280}, 1934--1937 (1998); S. Subramaniam, and R. Henderson 
Nature {\bf 406} 653--657 (2000); H.J. Sass {\it et al.} Nature {\bf 
406} 649--653 (2000); A. Royant {\it et al.} Nature {\bf 406} 645--648 (2000).

\bibitem{petitesves} M. Seigneuret and J-L. Rigaud,
FEBS Lett. {\bf 228}, 79--84 (1988);
M. Seigneuret and J-L. Rigaud,
Biochemistry {\bf 25}, 6723--6730 (1986);
J.-L. Rigaud, A. Bluzat, and S. B\"uschlen,
Biochem. Biophys. Res. Commun. {\bf 111}, 373--382 (1983).

\bibitem{Angelova} M. I. Angelova {\it et al.},
Prog. Coll. Polym. Science {\bf 89}, 127--131 (1992).

\bibitem{Daniel} D. L\'evy,
{\it personnal communication}.

\bibitem{Tocanne} F. Dumas {\it et al.},
Biophys. J. {\bf 73}, 1940--1953 (1997).

\bibitem{Fluo} J. Heberle and N. A. Dencher,
Proc. Nat. Acad. Sci. USA {\bf 89}, 5996-6000 (1992).

\bibitem{lightBR}
T. Kouyama, R. A. Bogomolni, and W. Stoeckenius,
Biophys. J. {\bf 48}, 201--208 (1985).

\bibitem{evans}
E. Evans and W. Rawicz,
Phys. Rev. Lett. {\bf 64}, 2094--2097 (1990);
E. Evans and D. Needham,
J. Phys. Chem. {\bf 91}, 4219 (1987).

\bibitem{ilnuovo}
W. Helfrich and R-M. Servuss, Il Nuovo
Cimento, {\bf 3D} 137--151 (1984).

\bibitem{EPC}
P. M{\'e}l{\'e}ard {\it et al.}, Biochimie {\bf 80}, 401--413 (1998);
G. Niggemann, N. Kummrow, and W. Helfrich,
J. Phys II France {\bf 5},413--425 (1995);
M. Kummrow and W. Helfrich, Phys. Rev. A {\bf 44}, 8356--8360 (1991);
F. Faucon {\it et al.}, J. Phys. France {\bf 50}, 2389--2414 (1989).

\bibitem{meleard97} Ph. M\'el\'eard {\it et al.},
Biophys. J. {\bf 72}, 2616--2629 (1997);
R. Kwok and E. Evans,
Biophys. J. {\bf 35}, 637--652 (1981).

\bibitem{effetglyc}
A. N. Radionov, and A. D. Kaulen,
FEBS Lett. {\bf 387}, 122--126 (1996);
Y. Cao {\it et al.}, Biochemistry {\bf 30}, 10972--10979 (1991).

\bibitem{trak}
F. Maingret, {\it et al.}, 
J. Biol. Chem. {\bf 274}, 1381--1387 and 26691--26696 (1999).

\bibitem{permeation}
M. Jansen and A. Blume, Biophys. J. {\bf 68}, 997--1008 (1995);
M. Jansen, {\it Thesis Universit\"at Keiserslautern} (1994);
R. Lawaczeck, Biophys. J. {\bf 45}, 491--494 (1984);
E. Boroske, M. Elwenspoek, and W. Helfrich, Biophys. J. {\bf 34},
95--109 (1981).

\bibitem{diffusion}
See for example :
O. G. Mouritsen and M. Bloom, Annu. Rev. Biophys. Biomol. Struct. {\bf 22},
145--171 (1993);
M. M. Speretto and O. G. Mouritsen, Biophys. J. {\bf 59},
261--271 (1991);
M. Bloom, E. Evans, and O. G. Mouritsen,
Q. Rev. Biophys. {\bf 24}, 293--297 (1991).






\bibitem{azide}
The Merck Index, Eleventh Edition (Merck \& Co, Inc.,
USA, Rackway, 1989) 1357.

\bibitem{petrov}
A.G. Petrov, Il Nuovo Cimento {\bf 3D}, 174--191 (1984).



\end{references}
\end{document}